\documentclass[conference]{IEEEtran}

\usepackage{amsmath}
\usepackage{amssymb}
\usepackage{amsfonts}
\usepackage{algorithmic}
\usepackage{graphicx}
\usepackage[dvipsnames]{xcolor}
\usepackage[free-standing-units,binary-units,per-mode=symbol,mode=text]{siunitx}
\usepackage{pgfplots}
\usepackage{tikz}
\usepackage{glossaries}
\usepackage{import}
\usepackage{microtype}
\usepackage{url}
\usepackage{enumitem}
\usepackage{multirow}
\usepackage{booktabs}
\usepackage[capitalise,noabbrev,nameinlink]{cleveref}
\usepackage{pgfplotstable}

\makeatletter
\let\MYcaption\@makecaption
\makeatother

\usepackage[font=footnotesize,position=b]{subcaption}

\makeatletter
\let\@makecaption\MYcaption
\makeatother

\pgfplotsset{compat=1.13}

\definecolor{color1}{HTML}{1269B0}
\definecolor{color2}{HTML}{910569}
\definecolor{color3}{HTML}{1DB24B}
\definecolor{color4}{HTML}{A8322C}
\definecolor{color5}{HTML}{F29545}

\DeclareSIUnit{\nothing}{\relax}
\DeclareSIUnit{\gateequi}{GE}

\newcommand\eg{e.g.,\ }
\newcommand\ie{i.e.,\ }

\newcommand{\SIadj}[2]{\SI[number-unit-product={\text{-}}]{#1}{#2}}
\newcommand{\FP}[1]{FP$_{\text{\textsc{#1}}}$}
\newcommand{\FPone}{\FP{1-sided}}
\newcommand{\FPtwo}{\FP{2-sided}}
\newcommand{\FPush}{\FP{u-shape}}
\newcommand{\aspectratio}[2]{\num{#1}:\num{#2}}
\newcommand{\commentout}[1]{}

\newacronym[longplural={Scratchpad Memories}]{SPM}{SPM}{Scratchpad Memory}
\newacronym{ACAP}{ACAP}{Adaptive Compute Acceleration Platform}
\newacronym{ACE}{ACE}{AXI Coherent Extensions}
\newacronym{AMBA}{AMBA}{Advanced Microcontroller Bus Architecture}
\newacronym{APB}{APB}{Advanced Peripheral Bus}
\newacronym{API}{API}{Application Programming Interface}
\newacronym{ASIC}{ASIC}{Application-Specific Integrated Circuit}
\newacronym{AVX}{AVX}{Advanced Vector Extension}
\newacronym{AXI}{AXI}{Advanced eXtensible Interface}
\newacronym{BLAS}{BLAS}{Basic Linear Algebra Subprograms}
\newacronym{CHI}{CHI}{Coherent Hub Interface}
\newacronym{CMOS}{CMOS}{Complementary Metal-Oxide-Semiconductor}
\newacronym{CNN}{CNN}{Convolutional Neural Network}
\newacronym{CPU}{CPU}{Central Processing Unit}
\newacronym{CSR}{CSR}{Control and State Register}
\newacronym{CTS}{CTS}{Clock Tree Synthesis}
\newacronym{DLP}{DLP}{Data Level Parallelism}
\newacronym{DMA}{DMA}{Direct Memory Access}
\newacronym{DRAM}{DRAM}{Dynamic Random-Access Memory}
\newacronym{DSP}{DSP}{Digital Signal Processing}
\newacronym{DUT}{DUT}{Device Under Test}
\newacronym{ECL}{ECL}{Emitter-Coupled Logic}
\newacronym{FBB}{FBB}{Forward Body-Biasing}
\newacronym{FDSOI}{FD-SOI}{Fully Depleted Silicon on Insulator}
\newacronym{FMA}{FMA}{Fused Multiply-Add}
\newacronym{FPGA}{FPGA}{Field-Programmable Gate Array}
\newacronym{FPU}{FPU}{Floating Point Unit}
\newacronym{GPGPU}{GPGPU}{General-Purpose \acrlong{GPU}}
\newacronym{GPU}{GPU}{Graphics Processing Unit}
\newacronym{HDL}{HDL}{Hardware Description Language}
\newacronym{HERO}{HERO}{Heterogeneous Embedded Research Platform}
\newacronym{HPC}{HPC}{High-Performance Computing}
\newacronym{ILP}{ILP}{Instruction Level Parallelism}
\newacronym{IOT}{IoT}{Internet-of-Things}
\newacronym{IP}{IP}{Intellectual Property}
\newacronym{IPC}{IPC}{Instructions Per Cycle}
\newacronym{IPU}{IPU}{Image Processing Unit}
\newacronym{ISA}{ISA}{Instruction Set Architecture}
\newacronym{LSU}{LSU}{Load/Store Unit}
\newacronym{LVT}{LVT}{low voltage threshold}
\newacronym{MIMD}{MIMD}{multiple instruction, multiple data}
\newacronym{MMU}{MMU}{Memory Management Unit}
\newacronym{MUL}{MUL}{multiplier}
\newacronym{MVL}{MVL}{maximum vector length}
\newacronym{NUMA}{NUMA}{non-uniform memory access}
\newacronym{NOC}{NoC}{Network-on-Chip}
\newacronym{PCIe}{PCIe}{Peripheral Component Interconnect Express}
\newacronym{PC}{PC}{Program Counter}
\newacronym{PE}{PE}{processing element}
\newacronym{PL}{PL}{Programmable Logic}
\newacronym{PMCA}{PMCA}{Programmable Manycore Accelerator}
\newacronym{PSL}{PSL}{Power Service Layer}
\newacronym{PTE}{PTE}{page-table entry}
\newacronym{PTW}{PTW}{page-table walker}
\newacronym{PULP}{PULP}{Parallel Ultra Low Power}
\newacronym{RAW}{RAW}{read-after-write}
\newacronym{RBB}{RBB}{Reverse Body-Biasing}
\newacronym{ROB}{ROB}{Reorder Buffer}
\newacronym{RTL}{RTL}{Register Transfer Level}
\newacronym{RVT}{RVT}{Regular Voltage Threshold}
\newacronym{RoCC}{RoCC}{Rocket Custom Coprocessor Interface}
\newacronym{SCM}{SCM}{Storage Class Memory}
\newacronym{SIMD}{SIMD}{single instruction, multiple data}
\newacronym{SIMT}{SIMT}{single instruction, multiple thread}
\newacronym{SLDU}{SLDU}{Slide Unit}
\newacronym{SLVT}{SLVT}{super-low voltage threshold}
\newacronym{SM}{SM}{Streaming Multiprocessor}
\newacronym{SRAM}{SRAM}{Static Random-Access Memory}
\newacronym{SSE}{SSE}{Streaming SIMD Extension}
\newacronym{SVE}{SVE}{Scalable Vector Extension}
\newacronym{TLP}{TLP}{Thread Level Parallelism}
\newacronym{TxnID}{TxnID}{Transaction ID}
\newacronym{VAC}{VAC}{Vector Access}
\newacronym{VC}{VC}{virtual channel}
\newacronym{VCONV}{VCONV}{Vector Conversion}
\newacronym{VEX}{VEX}{Vector Execute}
\newacronym{VFU}{VFU}{vector functional unit}
\newacronym{VID}{VID}{Vector Instruction Decode}
\newacronym{VIS}{VISSUE}{Vector Instruction Issue}
\newacronym{VLIW}{VLIW}{Very Long Instruction Word}
\newacronym{VLOOP}{VLOOP}{Vector Loop}
\newacronym{VLR}{VLR}{vector length register}
\newacronym{VLSU}{VLSU}{Vector Load/Store Unit}
\newacronym{VNB}{VNB}{Von Neumann Bottleneck}
\newacronym{VRF}{VRF}{Vector Register File}
\newacronym{VT}{VT}{vector thread}
\newacronym{WAR}{WAR}{write-after-read}
\newacronym{WAW}{WAW}{write-after-write}
\newacronym{DCT}{DCT}{discrete cosine transform}
\newacronym{QOR}{QoR}{quality of results}
\newacronym[firstplural=core complexes (CCs)]{CC}{CC}{core complex}
\newacronym{IDol}{I\$}{instruction cache}
\newacronym{HBM}{HBM}{High Bandwidth Memory}
\newacronym{HBI}{HBI}{High Bandwidth Interconnect}
\newacronym{CABF}{CABF}{Cull-and-Aggregate Bottom-Up Floorplanner}
\newacronym{FEOL}{FEOL}{front end of the line}
\newacronym{BEOL}{BEOL}{back end of the line}
\newacronym{F2F}{F2F}{face-to-face}
\newacronym{IMP}{IMP}{Iterative Merging Packing}
\newacronym{GPC}{GPC}{GPU Processing Cluster}
\newacronym{TPC}{TPC}{Texture Processing Cluster}
\newacronym{SoC}{SoC}{System-on-Chip}
\newacronym{PnR}{PnR}{Place and Route}
\newacronym{TNS}{TNS}{Total Negative Slack}
\newacronym{DP}{DP}{Datapath}
\newacronym{MulDiv}{MulDiv}{integer Multiply Divide Unit}
\newacronym{RtWL}{RtWL}{routed wirelength}
\newacronym{CGM}{CGM}{Core Memory Group}
\begin{document}

\title{Soft Tiles: Capturing Physical Implementation Flexibility for Tightly-Coupled Parallel Processing Clusters}

\author{\IEEEauthorblockN{Gianna Paulin\IEEEauthorrefmark{1}, %
Matheus Cavalcante\IEEEauthorrefmark{1}, %
Paul Scheffler\IEEEauthorrefmark{1}, %
Luca Bertaccini\IEEEauthorrefmark{1}, %
Yichao Zhang\IEEEauthorrefmark{1}, \\
Frank G\"urkaynak\IEEEauthorrefmark{1} and Luca Benini\IEEEauthorrefmark{1}\IEEEauthorrefmark{2}}\\
\IEEEauthorblockA{\IEEEauthorrefmark{1}Integrated Systems Laboratory, ETH Zurich\\
8092 Zurich, Switzerland\\
Email: \{pauling,matheus,paulsc,lbertaccini,yiczhang,kgf,lbenini\}@iis.ee.ethz.ch}
\IEEEauthorblockA{\IEEEauthorrefmark{2}University of Bologna\\
40126 Bologna, Italy}}

\maketitle

\begin{abstract}
Modern high-performance computing architectures (Multicore, GPU, Manycore) are based on tightly-coupled clusters of processing elements, physically implemented as rectangular tiles. Their size and aspect ratio strongly impact the achievable operating frequency and energy efficiency, but they should be as flexible as possible to achieve a high utilization for the top-level die floorplan. In this paper, we explore the flexibility range for a high-performance cluster of RISC-V cores with shared L1 memory used to build scalable accelerators, with the goal of establishing a hierarchical implementation methodology where clusters can be modeled as soft tiles to achieve optimal die utilization.
\end{abstract}

\begin{IEEEkeywords}
Floorplanning, Soft Blocks, VLSI Architectures.
\end{IEEEkeywords}

%%%%%%%%%%%%%%%%%%%%%%%%%%%%%%%%%%%%%%%%%%%%%%%%%%%%%%%%%%%%%%%%%%%%%%%%%
%%%  Introduction
%%%%%%%%%%%%%%%%%%%%%%%%%%%%%%%%%%%%%%%%%%%%%%%%%%%%%%%%%%%%%%%%%%%%%%%%%
\section{Introduction and Related Work}
\label{sec:introduction}

Floorplanning, the process of designing the physical layout of a chip, has a big impact on the performance, energy efficiency, time-to-market, and fabrication cost of VLSI chips.
While the main objective of floorplanning has been chip area reduction---which directly translates into lowering production costs---the floorplanning process must also optimize wirelength, delays, thermal stability, and energy efficiency~\cite{hoo2015cost}.
With the decreasing feature size of advanced nodes, the overall number of transistors per chip has skyrocketed.
Higher transistor densities have enabled larger chip designs, directly increasing the turnaround time of na\"ive iterative floorplan refinement.
To counteract this trend, researchers have explored new paradigms to accelerate the floorplanning process, leveraging \glspl{GPU}~\cite{lin2020dreamplace} and artificial intelligence~\cite{khailany2020accelerating}.

Despite such advanced paradigms, the high cell counts of today’s high-performance chips make a hierarchical implementation flow a necessity. They can be tackled by following a top-down or a bottom-up approach.
Top-down flows start by partitioning the die and allocating subregions of the chip layout to specific sub-blocks, generating constraints on their dimensions and aspect ratios.
When implementing the sub-blocks, such requirements might be unfeasible or lead to sub-optimal \gls{QOR}, requiring lengthy iterations to converge to a  feasible design.
A bottom-up flow would start by fully implementing and hardening blocks at lower hierarchy levels before implementing the next hierarchy level based on these tiles.
However, building upon hard tiles---which have fixed dimensions and even a fixed orientation in advanced technology nodes---might result in sub-optimal overall placement results, as they limit the possible top-level floorplans and make it more challenging to achieve a high utilization of the die area.

External \gls{IP} providers often deliver their modules as placed-and-routed hard tiles.
However, for \glspl{IP} owned by the chip designer, soft tiles are usually a better choice as their density and aspect ratios can be refined during placement.
Additionally, introducing soft tiles enables a broader range of possible floorplans with higher overall utilization, especially when the die dimensions are fixed (fixed-outline floorplanning) and the designer needs to fulfill stringent requirements such as tight frequency and power constraints~\cite{adya2003fixed,chen2009floorplanning}.
Various works have investigated algorithms tackling the floorplanning challenges for designs with hard and soft tiles, \eg \gls{CABF}~\cite{hoo2015cost} or \gls{IMP}~\cite{ji2017iterative}.
% Various floorplanning algorithms such as \paulsc{the} \gls{CABF}~\cite{hoo2015cost} or \gls{IMP}~\cite{ji2017iterative} can tackle the additional challenges of not only using hard tiles, but also soft tiles in hierarchical floorplans.
However, the correlation between a soft tile's physical shape (\eg aspect ratio and macro placement) in floorplanning and its \gls{QOR} after physical implementation has barely been analyzed so far.

%However, the correlation between floorplan decisions (e.g., aspect ratio, macro placement, and cell density) and the QoR of the soft tiles has not been analyzed, despite their increasing importance for modern high-performance systems.
Typically, modern architectures build upon a base compute cluster, combining multiple \glspl{PE} sharing access to an L1 cache or \gls{SPM} via a low-latency interconnect. We call such a latency-critical cluster a tile. This tile is then replicated and interconnected with a latency-tolerant \gls{NOC} to build a larger high-performance system.
For example, considering multicore processors, Fujitsu’s A64FX~\cite{yoshida2018fujitsu} combines four interconnected \glspl{CGM}.
Each \gls{CGM} couples twelve compute (and one control) superscalar out-of-order vector-capable cores with a fast data cache.
In addition, each \gls{CGM} has a designated \gls{HBM} controller enabling a bandwidth of \SI{256}{\giga\byte/\second} to \SI{8}{\gibi\byte} of \gls{HBM}.
As an example in the \gls{GPU} field, NVIDIA’s A100~\cite{choquette2021nvidia} architecture contains 108~\glspl{SM} grouped into Texture Processing Clusters, which in turn are grouped into \gls{GPU} Processing Clusters.
Each \gls{SM} has a combined L1 data cache, a shared memory of \SI{192}{\kibi\byte}, and four warps. Each warp includes a tensor core, 16 INT32 and FP32 cores, eight FP64 cores, and a large shared \SI{64}{\kibi\byte} register file.
Finally, Esperanto's ET-SoC-1~\cite{ditzel2021accelerating}, an exemplary manycore accelerator, couples more than one thousand energy-efficient RISC-V vector processors, each including a software-configurable L1 data cache or \gls{SPM}, with four high-performance Linux-capable out-of-order \glspl{CPU}. Esperanto uses a highly regular tiled architecture to fit all these cores on a chip. Eight energy-efficient RISC-V processors with a \SI{32}{\kibi\byte} shared \gls{IDol} form a \textit{neighborhood}, and four neighborhoods with a \SI{4}{\mebi\byte} L2 \gls{SRAM} form a \textit{minion shire}. Finally, 34 minion shires with \SI{136}{\mebi\byte} of on-die memory and four Linux-capable \glspl{CPU} form the full ET-SoC chip.

Finding an optimal floorplan with good post-place-and-route \gls{QOR} for the tightly interconnected tile (multicore, GPU, manycore) is crucial for achieving high performance and efficiency of the overall design. The tile's operating frequency ultimately determines the performance of the overall architecture.
Moreover, the L1-to-\gls{PE} latency is critical for high \gls{IPC} and cannot easily be increased by pipelining to simplify implementation.
In this scenario, the soft tile and system-level \gls{QOR} need to be improved through physically-aware design approaches~\cite{cortadella2013physical, cavalcante2021mempool}.

This paper focuses on an open-source, high-performance cluster tile with eight compute (and one control) RISC-V cores connected to a shared L1 \gls{SPM} through a low-latency interconnect~\cite{zaruba2020snitch}.
Similarly to the state-of-the-art architectures, the cluster tile is then replicated to build a scaled-up high-performance acceleration system~\cite{zaruba2021}.
We explore the \gls{QOR} of the physical implementation of this cluster as a soft tile based on a flexible range of aspect ratio and memory macro placement styles.
The contributions of this paper are:
\begin{itemize}
    \item We propose three different memory macro placement styles for a cluster of RISC-V cores with a shared L1 \gls{SPM}. For each, we place and route the cluster and evaluate its \gls{QOR} in terms of achievable operating frequency, \gls{TNS}, number of violating paths, cell density, total \gls{RtWL}, and number of inserted buffers.
    \item For all three proposed placement styles, we explore and evaluate the impact of the tile's aspect ratio on the \gls{QOR}.
    \item Based on our results, we discuss a hierarchical implementation methodology where clusters can be modeled as soft tiles to achieve optimal overall die utilization.
\end{itemize}

\section{Architecture}

The Snitch cluster~\cite{zaruba2020snitch} is an open-source RISC-V multicore cluster targeting highly-efficient double-precision floating-point computing. It is the compute unit of the Manticore architecture~\cite{zaruba2021} where it is massively replicated.

\begin{figure}[t]
  \centering
  \includegraphics[width=\linewidth]{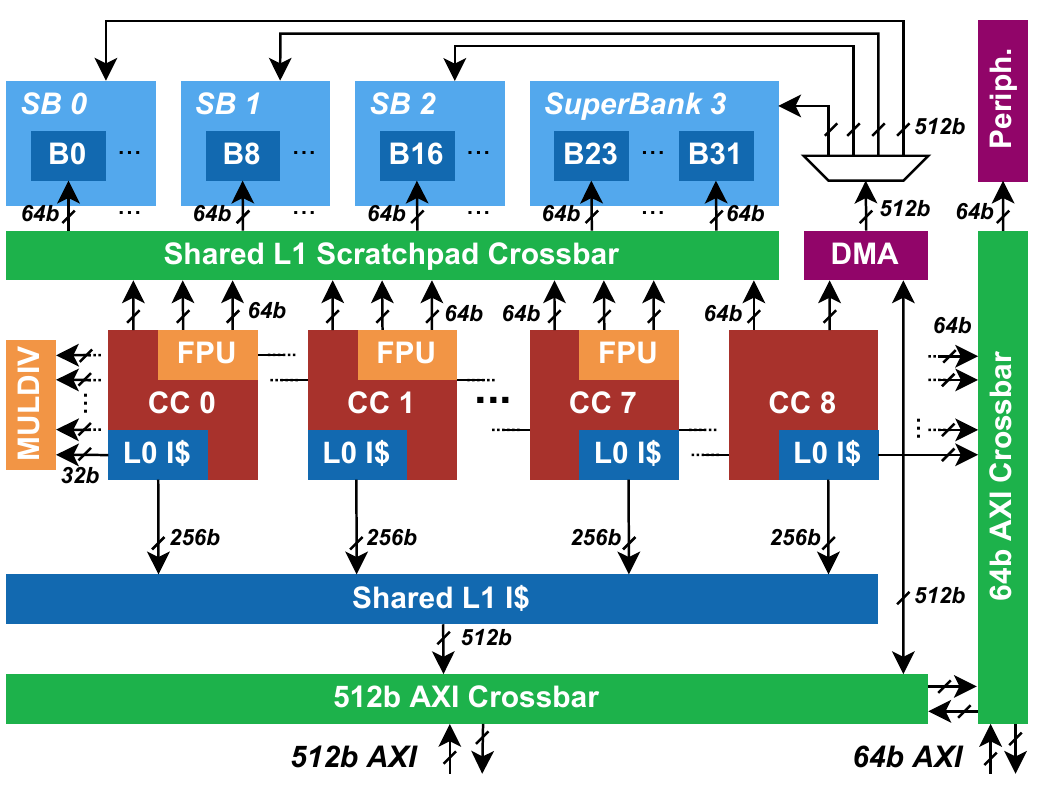}
  \caption{Architecture of a Manticore Snitch cluster.}
    \label{fig:arch_cluster}
\end{figure}

\subsection{Cluster Tile}
\label{subsec:arch_cluster}

% Working on the figure RN
% Miniature version of Occamy cluster figure
\cref{fig:arch_cluster} shows the architecture of the Snitch cluster configuration used in Manticore. It contains eight worker \glspl{CC}, each of which combines a small integer core, a trace L0 instruction cache, and a large double-precision \gls{FPU} kept busy with custom architectural extensions. An additional ninth \gls{CC} without an \gls{FPU} controls a cluster-level \gls{DMA} engine and can be used for cluster coordination.
All \glspl{CC} share a tightly-coupled \SI{128}{\kibi\byte} L1 \gls{SPM} divided into 32 memory banks, each \SI{64}{\bit} wide, via a single-cycle \gls{SPM} interconnect. Blocks of eight banks form \emph{superbanks}, which are accessed in parallel by the \SIadj{512}{\bit} \gls{DMA} engine through a secondary wide interconnect. The \glspl{CC} also share a two-way \SI{8}{\kibi\byte} L1 \gls{IDol} and an \gls{MulDiv}.
Finally, the \gls{DMA} engine and L1 \gls{IDol} share a duplex \SIadj{512}{\bit} \gls{AXI} crossbar connection to the global memory system, which all \glspl{CC} can access through a \SIadj{64}{\bit} secondary \gls{AXI} crossbar; both crossbars are internally connected for convenience.

\begin{figure*}[t]
  \begin{minipage}{0.32\linewidth}
    \centering
    \includegraphics[width=\linewidth]{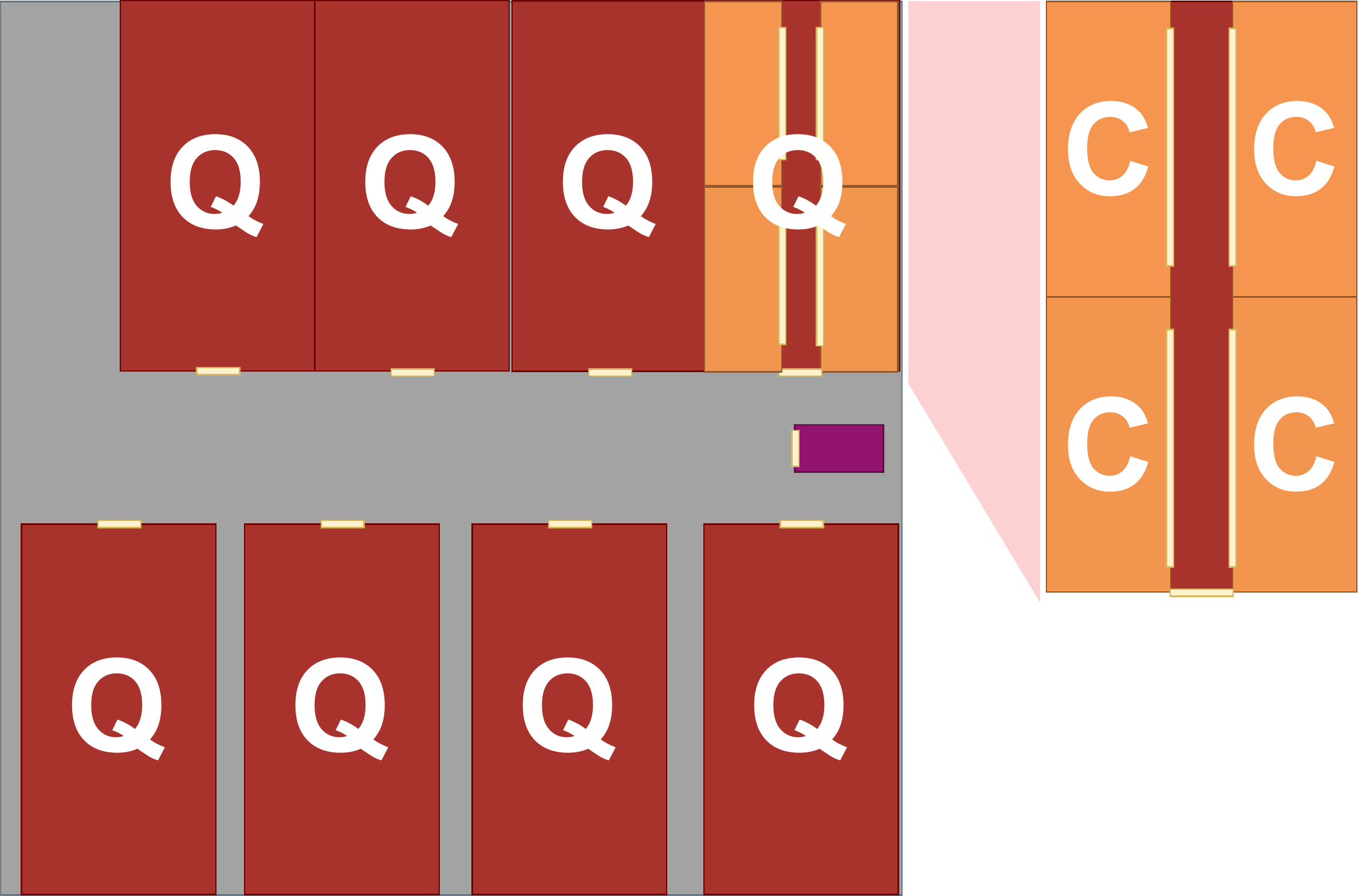}
    \subcaption{Wide cluster tile.}
    \label{fig:fp_onesided}
  \end{minipage}\hfill%
  \begin{minipage}{0.34\linewidth}
    \centering
    \includegraphics[width=\linewidth]{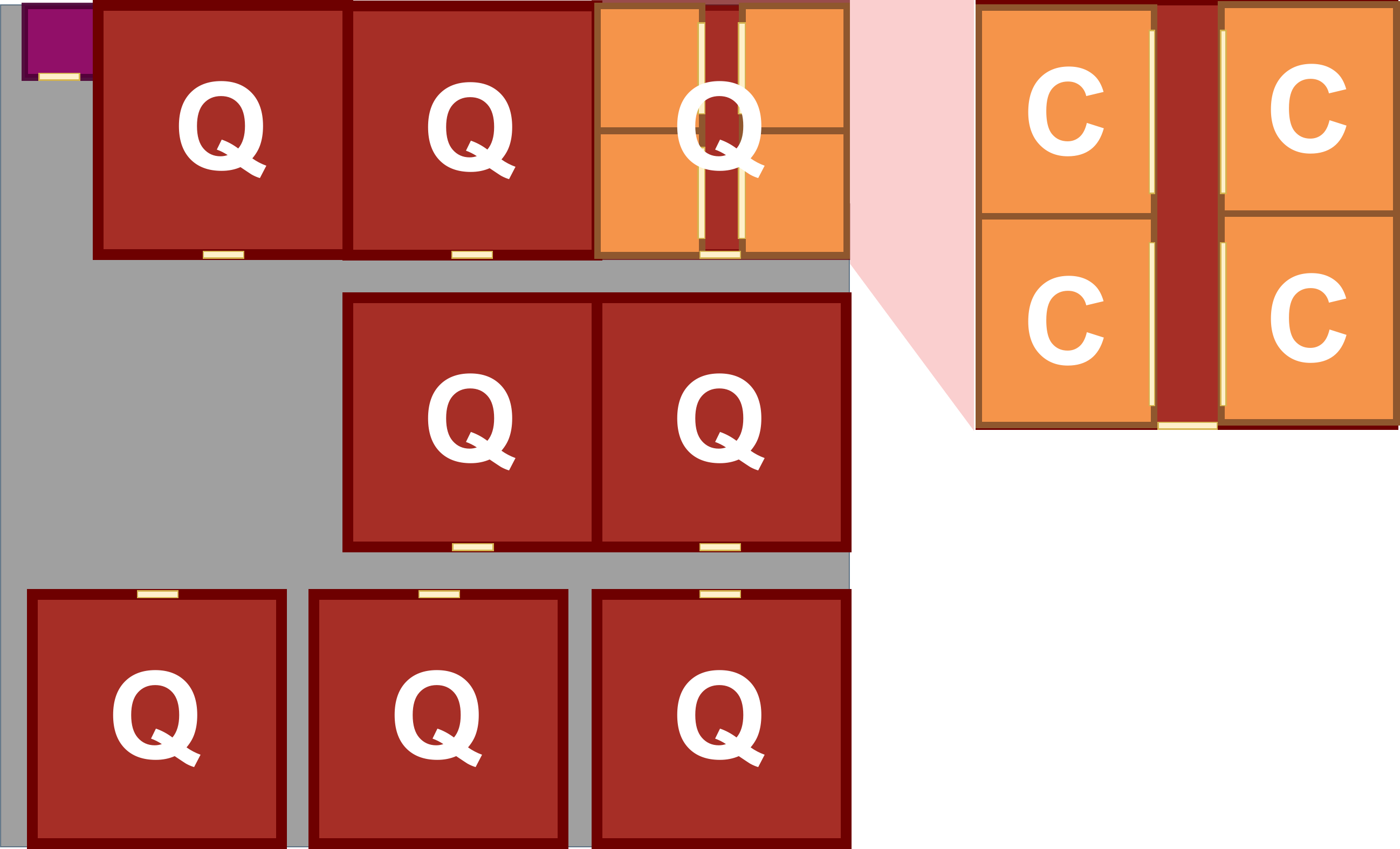}
    \subcaption{Square cluster tile.}
    \label{fig:fp_twosided}
  \end{minipage}\hfill%
%   \vspace{.8em}
  \begin{minipage}{0.32\linewidth}
    \centering
    \includegraphics[width=\linewidth]{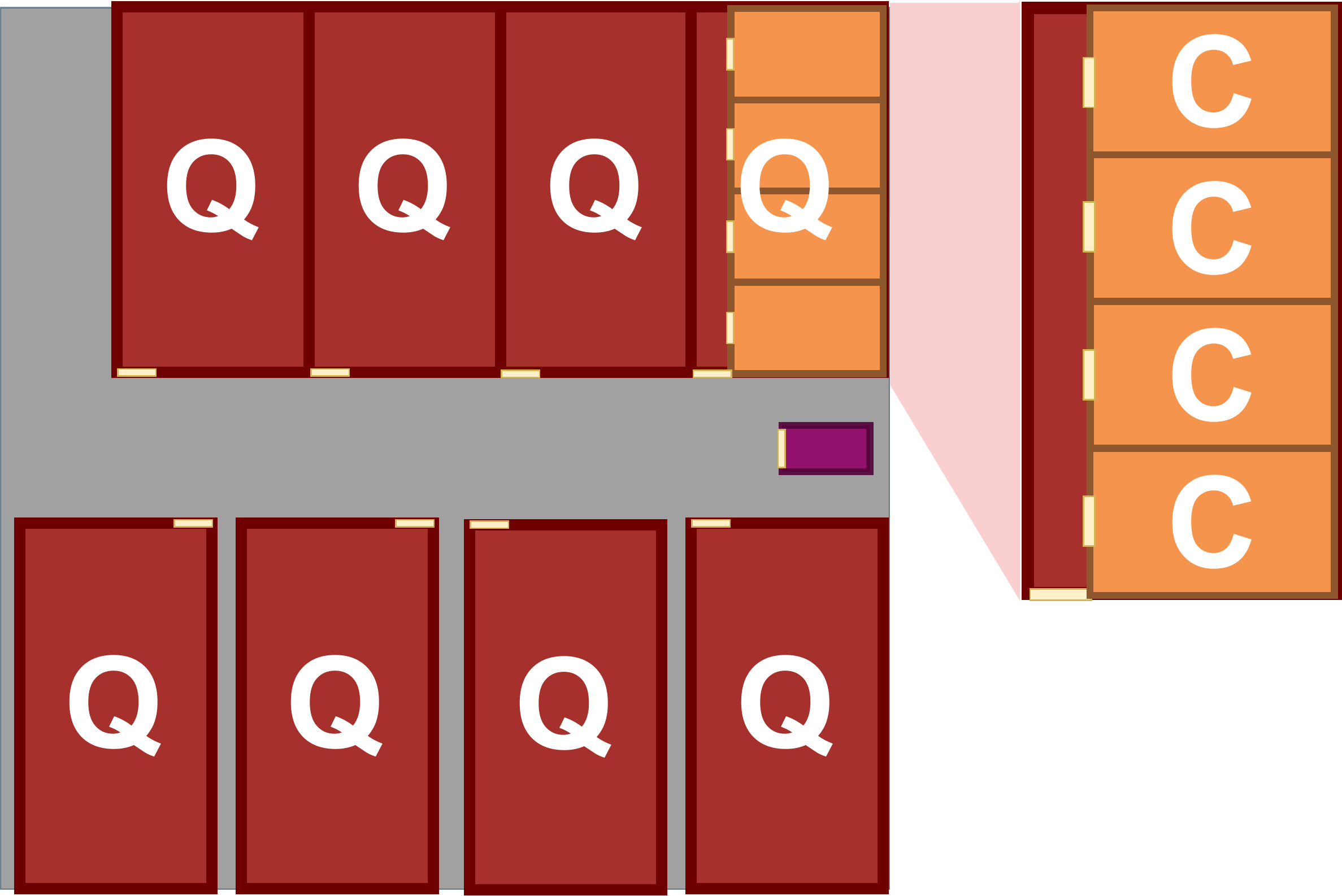}
    \subcaption{Tall cluster tile.}
    \label{fig:fp_ushape}
  \end{minipage}

  \caption{Three example top-level floorplans containing eight quadrants $\mathcal{Q}$ with 4 clusters $\mathcal{C}$ each and one manager core (purple). I/O ports are highlighted in yellow.}
  \label{fig:fp_toplevel}
\end{figure*}

% I thought we'd combine the quadrant and top into one section, since this is existing work anyway and not the focus
\subsection{System Integration}
\label{subsec:sys_integration}

The cluster tile can be hierarchically replicated to form a manycore system with thousands of cores~\cite{zaruba2021}. For example, four clusters $\mathcal{C}$ can be combined to form a quadrant $\mathcal{Q}$, an intermediate hierarchy level with a shared read-only cache and connections to the memory system. Multiple quadrants can then be combined to form the top-level manycore architecture, which also includes application-grade manager cores, high-bandwidth die-to-die interfaces, and additional peripherals.

Unlike the cluster tile with its tightly-coupled low-latency memories, the global interconnect uses pipelineable and latency-tolerant links; it is, therefore, less critical and can easily adapt to changing placement and routing pressures.
Thus, floorplanning efforts should be focused on the massively replicated compute tile, which is not latency tolerant and highly frequency-critical, therefore dictating the system's performance, area, and operating frequency.
Nevertheless, the hierarchical scale-out strategy can impose additional constraints onto the repeated tile, such as aspect ratio or pin positioning.
% The hierarchical scale-out strategy can impose additional constraints  on the floorplan of the repeated tile with each level.
\Cref{fig:fp_toplevel} shows three examples of how different quadrant and top-level organizations can require the cluster tile to take specific aspect ratios, further motivating our experiments.

%%%%%%%%%%%%%%%%%%%%%%%%%%%%%%%%%%%%%%%%%%%%%%%%%%%%%%%%%%%%%%%%%%%%%%%%%
%%%%%%%%%%%%%%%%%%%%%%%%%%%%%%%%%%%%%%%%%%%%%%%%%%%%%%%%%%%%%%%%%%%%%%%%%

%%%%%%%%%%%%%%%%%%
%  Methodology   %
%%%%%%%%%%%%%%%%%%

%%%%%%%%%%%%%%%%%%%%%%%%%%%%%%%%%%%%%%%%%%%%%%%%%%%%%%%%%%%%%%%%%%%%%%%%%
%%%%%%%%%%%%%%%%%%%%%%%%%%%%%%%%%%%%%%%%%%%%%%%%%%%%%%%%%%%%%%%%%%%%%%%%%

\section{Methodology}
\label{sec:methodology}

We use \textsc{Synopsys Fusion Compiler 2020.09} to synthesize, place, and route the Manticore Snitch cluster in \textsc{GlobalFoundries}' \SI{12}{\nano\meter} advanced FinFET technology node for a set of memory placement styles and aspect ratios ranging from very wide (\aspectratio{2.5}{1}) to very tall (\aspectratio{1}{2.5}). All shown designs target a \SI{1}{\giga\hertz} clock frequency under worst-case conditions (SS, \SI{0.72}{\volt}, \SI{125}{\celsius}) with a core area of \SI{0.90}{\square\milli\meter}. The designs were taken to the route optimization stage, with the tool trying to solve as many DRC violations as possible but without running a sign-off phase. Finally, we evaluate the physically implemented designs in terms of:
\begin{description}
    \item[Effective frequency] Limits the compute throughput per tile.
    \item[Cell density] Limits the placeable logic (tiles) per unit die area.
    \item[\#Buffers] Increases with expended timing and design rule fixing effort and impacts leakage and switching power.
    \item[\Acrfull{RtWL}] Worsens the transition times, crosstalk, and ohmic losses, requiring more timing and design rule fixing and increasing leakage and switching power.
    \item[\#DRC Violations] Indicates the routability of the design.
    \item[\#Violating Paths/\gls{TNS}] Capture the overall \emph{severity} and \emph{ubiquity} of timing violations and increase with the optimization effort required to achieve a given effective frequency.
\end{description}

%%%%%%%%%%%%%%%%%%%%%%%%%%%%%%%%%%%%%%%%%%%%%%%%%%%%%%%%%%%%%%%%%%%%%%%%%
%%%%%%%%%%%%%%%%%%%%%%%%%%%%%%%%%%%%%%%%%%%%%%%%%%%%%%%%%%%%%%%%%%%%%%%%%

%%%%%%%%%%%%%%%%%%%%%%%
%   Base Floorplans   %
%%%%%%%%%%%%%%%%%%%%%%%

%%%%%%%%%%%%%%%%%%%%%%%%%%%%%%%%%%%%%%%%%%%%%%%%%%%%%%%%%%%%%%%%%%%%%%%%%
%%%%%%%%%%%%%%%%%%%%%%%%%%%%%%%%%%%%%%%%%%%%%%%%%%%%%%%%%%%%%%%%%%%%%%%%%

\section{Floorplanning}
\label{sec:impl_basefp}

\begin{figure*}[t]
  \begin{minipage}{.30\textwidth}
    \centering
    \includegraphics[width=\textwidth]{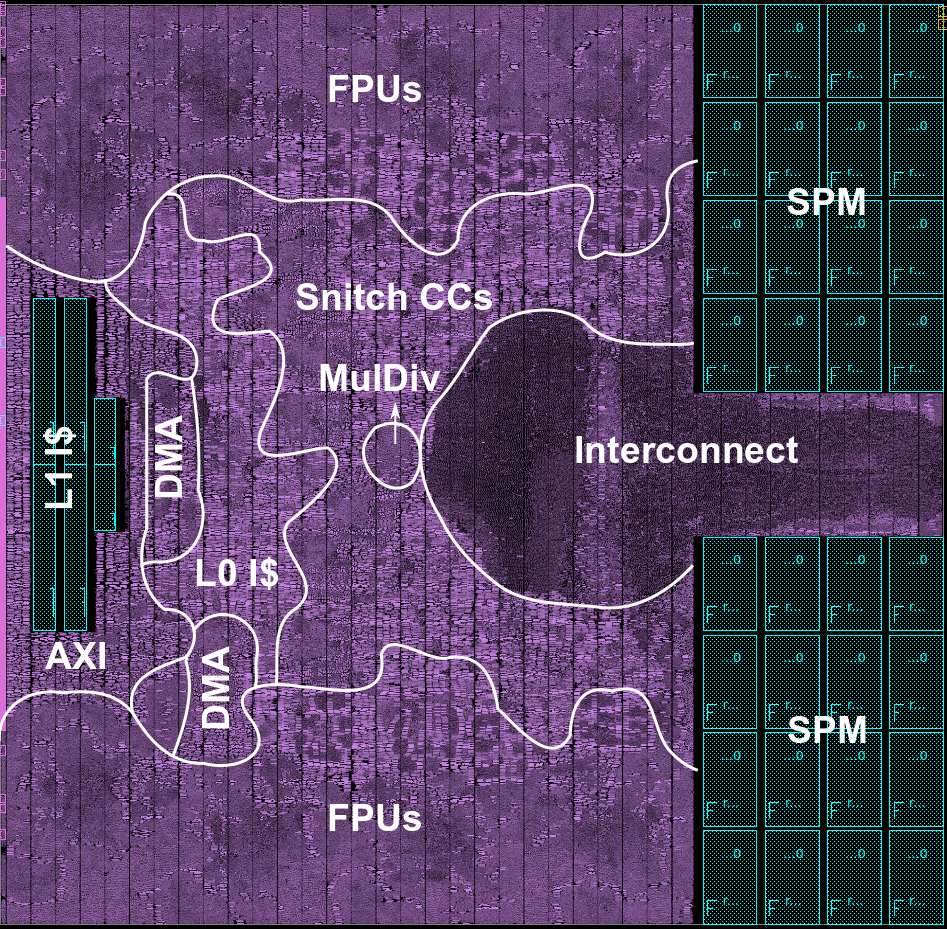}
    \subcaption{\FPone{}.}
    \label{fig:fp_base_onesided}
  \end{minipage}\hfill%
  \begin{minipage}{.30\textwidth}
    \centering
    \includegraphics[width=\textwidth]{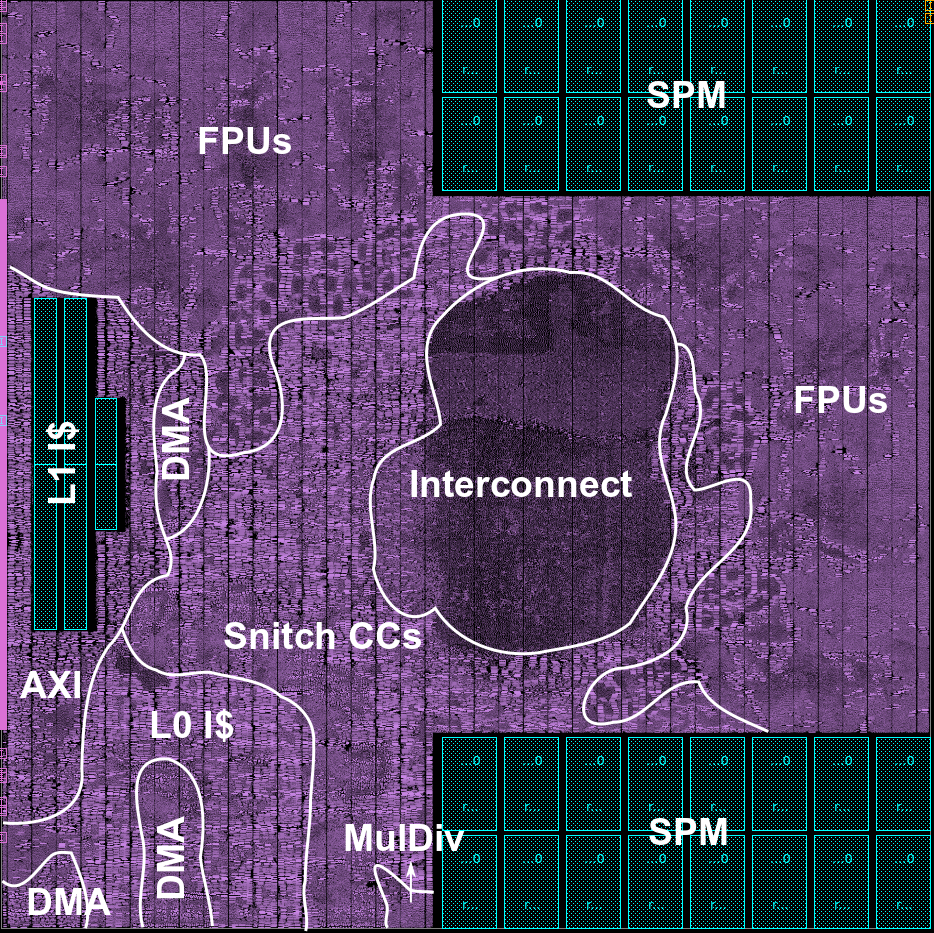}
    \subcaption{\FPtwo{}.}
    \label{fig:fp_base_twosided}
  \end{minipage}\hfill%
  \begin{minipage}{.30\textwidth}
    \centering
    \includegraphics[width=\textwidth]{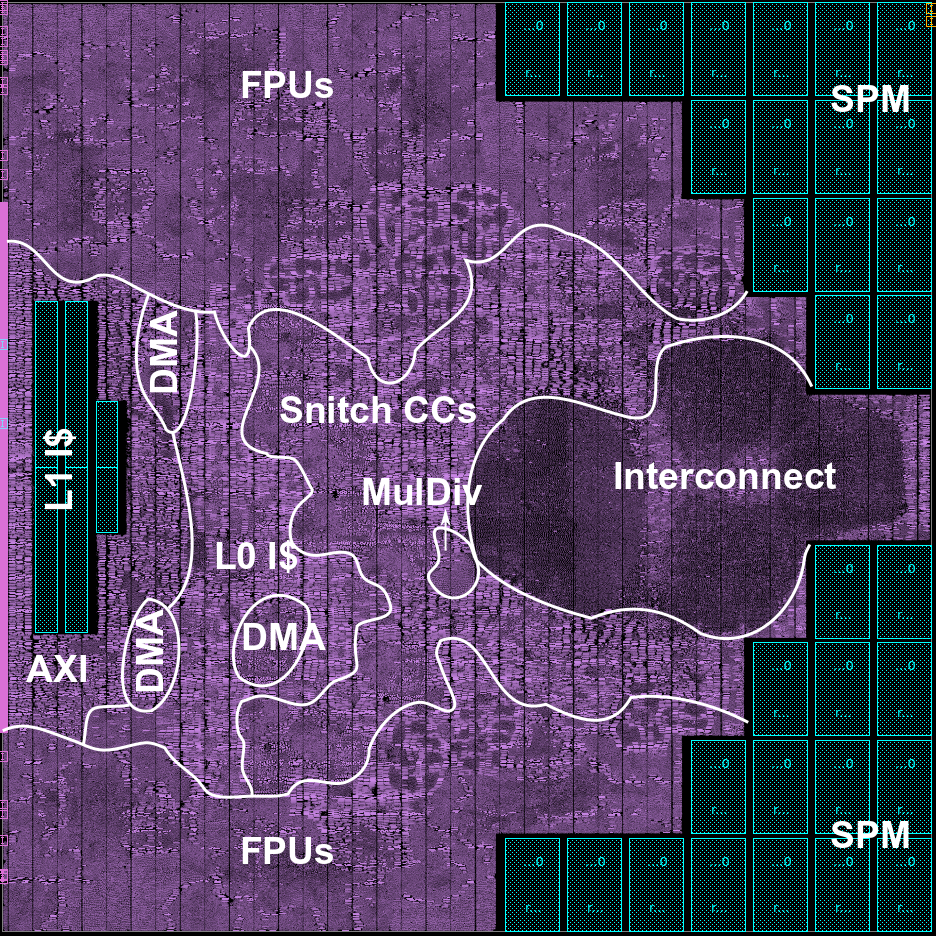}
    \subcaption{\FPush{}.}
    \label{fig:fp_base_ushape}
  \end{minipage}%
  \caption{Placed and routed designs for each base floorplan with main modules highlighted.}
  \label{fig:meth_basefps}
\end{figure*}

To arrive at our evaluation floorplans, we will first observe the cluster tile's architecture by analyzing the interconnectivity of individual design components since these properties will fundamentally impact our target metrics.

Technology-dependent SRAM macros are used to implement all \gls{IDol} and \gls{SPM} banks.
With 32 of 36 memory macros and \SI{20}{\percent} of the design's placeable area, the L1 \gls{SPM} is the most challenging component to place.
Its fully-connected crossbar dominates all other interconnect logic in complexity, providing an individual \SIadj{64}{\bit} low-latency link for all \num{26x32} master-slave pairs; we can predict that it will be routing-dominated and, therefore, susceptible to low cell density and routing congestion.
In addition, this interconnect requires a large unobstructed placement area and for all the \gls{SPM} macros to be close to it.

We propose the three representative floorplan styles to explore the \gls{QOR} of the Manticore Snitch cluster: \FPone{}, \FPtwo{}, and \FPush{}.
The I/O pins are constrained to the left side, and the \gls{SPM} macros are on the right side, in all floorplan styles.
The \gls{IDol} macros are placed close to the left side not to obstruct the \gls{SPM} crossbar yet remain accessible to the \glspl{CC}.
Moreover, all styles are vertically symmetric to keep the interconnect easily reachable by all \glspl{CC} and \gls{SPM} macros.

\textbf{\FPone{}}, shown in \Cref{fig:fp_base_onesided}, places all \num{32} \gls{SPM} memory macros on the right side in a block as compact as possible.
It tries to keep the interconnect as close as possible to the right side of the design.
This placement style might lead to narrow channels in the center of the design, causing the ``pinching'' of the interconnect area.
Moreover, the excessive macro stacking causes some macros to be far from the standard cell area, which challenges the routing to their pins.

\textbf{\FPtwo{}}, shown in \Cref{fig:fp_base_twosided}, tackles those limitations by leaving a largely unconstrained placement area for the interconnect and keeping all \gls{SPM} macros easily accessible from the standard cell area.
To do so, it places the \gls{SPM} macros in a block as wide as possible.
However, this placement style spreads the macros across the design, which might affect the wire length and the timing.

\textbf{\FPush{}}, shown in \Cref{fig:fp_base_ushape}, is a midway point between the two previous styles.
This floorplan places the \gls{SPM} macros in a vertically symmetrical ``U'' shape enclosing the region where the \gls{SPM} crossbar is placed; this further minimizes their overall distance of the \gls{SPM} crossbar and avoids excessive stacking.
The \gls{SPM} macros are placed according to samples of the generator function $f[n] = (n/\text{HH})^{-p}$, where $n \in \mathbb{Z}^*$ represents the $n$-th column of macros in the \gls{SPM} region, $f[n]$ represents the height of the corresponding column, $\text{HH}$ is half of the height of the cluster area, and the parameter $p$ is chosen so that the \gls{SPM} macros occupy as many rows as possible (\ie closing the central channel).

%%%%%%%%%%%%%%%%%%%%%%%%%%%%%%%%%%%%%%%%%%%%%%%%%%%%%%%%%%%%%%%%%%%%%%%%%
%%%%%%%%%%%%%%%%%%%%%%%%%%%%%%%%%%%%%%%%%%%%%%%%%%%%%%%%%%%%%%%%%%%%%%%%%

%%%%%%%%%%%%%%%
%   Results   %
%%%%%%%%%%%%%%%

%%%%%%%%%%%%%%%%%%%%%%%%%%%%%%%%%%%%%%%%%%%%%%%%%%%%%%%%%%%%%%%%%%%%%%%%%
%%%%%%%%%%%%%%%%%%%%%%%%%%%%%%%%%%%%%%%%%%%%%%%%%%%%%%%%%%%%%%%%%%%%%%%%%

\section{Implementation Results}
\label{sec:implementation}

This section compares the \gls{QOR} for our three different memory macro placement styles for a range of aspect ratios. %
We will then discuss how understanding the aspect ratio scaling of different tile floorplans can help determine an optimal hierarchical top-level floorplan.
% and evaluate a selection of floorplans on their \emph{compressibility}, that is, their ability to maintain their \gls{QOR} when their area is scaled down.

%%%%%%%%%%%%%%%%%%%%%%%%%%%%%%%%%%%%%%%%%%%%%%%%%%%%%%%%%%%%%%%%%%%%%%%%%
\subsection{Base Floorplan Analysis}
\label{ssec:base_fp}

\Cref{fig:fp_base_onesided,fig:fp_base_twosided,fig:fp_base_ushape} highlight the placement of the main modules of the Manticore Snitch cluster.
The interconnect occupies a large central area in the design, encircled by the \glspl{CC}.
The \glspl{CC} are attracted by the L1 \gls{IDol} and the \gls{SPM} crossbar, while the \glspl{FPU} occupy the remaining area.

The Snitch cluster's placeable cell area is approximately \SI{4.5}{\mega\gateequi}.
The uniform cluster area of \SI{0.9}{\square\milli\meter} implies an average standard cell utilization of \SI{57.5}{\percent} over the three base floorplans.
The standard cell density maps for all base floorplans are shown in \cref{fig:cell_density_oneside_Q1.0,fig:cell_density_twoside_Q1.0,fig:cell_density_ushape_Q1.0}.
The low cell density of the densely routed \gls{SPM} crossbar indicates that the density in the regions where the \glspl{CC} and \glspl{FPU} are placed is, in fact, much higher than the average density suggests, peaking at \SI{80}{\percent}.
This puts pressure on the rest of the design, which ends up dense and having to avoid the routing-congested interconnect region.

For \cref{fig:fp_base_onesided,fig:fp_base_ushape}, the modules were placed in suitable locations.
However, in \Cref{fig:fp_base_twosided}, the lack of space between the interconnect and the L1 \gls{IDol} macros for the \glspl{CC} and L0 \gls{IDol}es results in an asymmetric module placement, where the cores end up on the bottom of the floorplan.
Furthermore, the module positions indicate that this floorplan style leads to a cluster particularly sensitive to changes in the aspect ratio in the direction of taller (narrower) clusters since this will further reduce the distance between the crossbar and the L1 \gls{IDol} macros.

\Cref{tab:aspect_ratio_results} compares those three floorplan styles according to several \gls{QOR} metrics.
\FPush{} shows the best \gls{QOR} overall with a square \aspectratio{1}{1} aspect ratio, reaching the highest operating frequency, lowest \gls{TNS}, fewer violating paths, and shorter \gls{RtWL}.
\FPone{} reaches comparable results, the main difference being an operating frequency \SI{1.5}{\percent} lower than the \FPush{} instance.
This performance drop is due to pinching of the interconnect by the central channel in the \FPone{} instance.

Both the \FPone{} and \FPush{} instances finished with 38 DRC violations.
This low ($<100$) DRC violation count is expected, since we only run our implementation flow to the route optimization stage.
Those violations are spread across the floorplan and can be solved manually in a sign-off step.
\FPtwo{}, on the other hand, had 227 violations, many of them shorts concentrated in the interconnect region.
Therefore, we consider this instance unfeasible.
This high DRC count is due to the interconnect placement between the \glspl{FPU} and their \glspl{CC}.
There are not enough resources to route those connections through the interconnect region, which leads to a high amount of DRCs and an unfeasible design.

%%%%%%%%%%%%%%%%%%%%%%%%%%%%%%%%%%%%%%%%%%%%%%%%%%%%%%%%%%%%%%%%%%%%%%%%%
\subsection{Aspect Ratio Analysis}
\label{ssec:aspectratio}

\begin{figure}[t]
  \centering
  \begin{minipage}{\linewidth}
    % Adapt horizontal offset here
    \hspace{1.2cm}
    \begin{tikzpicture}
        \begin{axis}[%
          hide axis,
          xmin=0, xmax=1, ymin=0, ymax=1,
          legend style={
            legend columns=-1,
            % Adapt total legend width here
            column sep=0.125cm
          }
        ]
        \addlegendimage{very thick, color2,mark=square*}
        \addlegendentry{\FPone{}};
        \addlegendimage{very thick, color5,mark=diamond*}
        \addlegendentry{\FPtwo{}};
        \addlegendimage{very thick, color3,mark=triangle*}
        \addlegendentry{\FPush{}};
        \end{axis}
    \end{tikzpicture}
  \end{minipage}%
  \hfill\allowbreak%
  \begin{minipage}[t]{0.48\linewidth}
    \centering
    \resizebox{0.98\linewidth}{!}{\begin{tikzpicture}
        \begin{axis}[
            height=4.5cm,
            width=0.98\linewidth,
            title style={at={(0.5,-0.25)},anchor=north,yshift=0.0},
            ylabel={Eff. Freq. [\si{\mega\hertz}]},
            ymin=860,
            ymax=960,
            ymajorgrids=true,
            yminorgrids=true,
            xlabel={Aspect Ratio},
            xtick=\empty,
            xmajorgrids=true,
            extra x ticks={0, 1, 2},
            extra x tick labels={\aspectratio{1}{2.5}, \aspectratio{1}{1}, \aspectratio{2.5}{1}}]

            % Plot the results
            \addplot [very thick, mark=square*,   color2] table [x=Index, y=Frequency] {results/FPone.txt};
            \addplot [very thick, mark=diamond*,  color5] table [x=Index, y=Frequency] {results/FPtwo.txt};
            \addplot [very thick, mark=triangle*, color3] table [x=Index, y=Frequency] {results/FPush.txt};
        \end{axis}
    \end{tikzpicture}}%\hfill%
    \phantomsubcaption{}
    \label{fig:timing_vs_aspect_ratio0}
  \end{minipage}%
  \hfill\allowbreak%
  \begin{minipage}[t]{0.48\linewidth}
    \centering
    \resizebox{0.98\linewidth}{!}{\begin{tikzpicture}
        \begin{axis}[
            height=4.5cm,
            title style={at={(0.5,-0.25)},anchor=north,yshift=0.0},
            width=0.98\linewidth,
            ylabel={Routed Wirel. [\si{\meter}]},
            ymin=15,
            ymax=18,
            ymajorgrids=true,
            yminorgrids=true,
            xlabel={Aspect Ratio},
            xtick=\empty,
            xmajorgrids=true,
            extra x ticks={0, 1, 2},
            extra x tick labels={\aspectratio{1}{2.5}, \aspectratio{1}{1}, \aspectratio{2.5}{1}}]

            % Plot the results
            \addplot [very thick, mark=square*,   color2] table [x=Index, y=RtWL] {results/FPone.txt};
            \addplot [very thick, mark=diamond*,  color5] table [x=Index, y=RtWL] {results/FPtwo.txt};
            \addplot [very thick, mark=triangle*, color3] table [x=Index, y=RtWL] {results/FPush.txt};
        \end{axis}
    \end{tikzpicture}}
    \phantomsubcaption{}
    \label{fig:wirelength_vs_aspect_ratio1}
    \end{minipage}%
    \hfill\allowbreak%
    \begin{minipage}[t]{0.48\linewidth}
    \centering
    \resizebox{0.98\linewidth}{!}{\begin{tikzpicture}
        \begin{axis}[
            height=4.5cm,
            width=0.98\linewidth,
            title style={at={(0.5,-0.25)},anchor=north,yshift=0.0},
            ylabel={\# DRCs [$\times 10^3$]},
            ymin=0,
            ymax=3,
            ymajorgrids=true,
            yminorgrids=true,
            xlabel={Aspect Ratio},
            xtick=\empty,
            xmajorgrids=true,
            extra x ticks={0, 1, 2},
            extra x tick labels={\aspectratio{1}{2.5}, \aspectratio{1}{1}, \aspectratio{2.5}{1}}]

            % Plot the results
            \addplot [very thick, mark=square*,   color2] table [x=Index, y=DRCs] {results/FPone.txt};
            \addplot [very thick, mark=diamond*,  color5] table [x=Index, y=DRCs] {results/FPtwo.txt};
            \addplot [very thick, mark=triangle*, color3] table [x=Index, y=DRCs] {results/FPush.txt};
        \end{axis}
    \end{tikzpicture}}%\hfill%
    \phantomsubcaption{}
    \label{fig:drc_vs_aspect_ratio0}
  \end{minipage}%
  \hfill\allowbreak%
  \begin{minipage}[t]{0.48\linewidth}
    \centering
    \resizebox{0.98\linewidth}{!}{\begin{tikzpicture}
        \begin{axis}[
            height=4.5cm,
            title style={at={(0.5,-0.25)},anchor=north,yshift=0.0},
            width=0.98\linewidth,
            ylabel={\#Buffers. [$\times 10^3$]},
            ymin=120,
            ymax=150,
            ymajorgrids=true,
            yminorgrids=true,
            xlabel={Aspect Ratio},
            xtick=\empty,
            xmajorgrids=true,
            extra x ticks={0, 1, 2},
            extra x tick labels={\aspectratio{1}{2.5}, \aspectratio{1}{1}, \aspectratio{2.5}{1}}]

            % Plot the results
            \addplot [very thick, mark=square*,   color2] table [x=Index, y=Buffers] {results/FPone.txt};
            \addplot [very thick, mark=diamond*,  color5] table [x=Index, y=Buffers] {results/FPtwo.txt};
            \addplot [very thick, mark=triangle*, color3] table [x=Index, y=Buffers] {results/FPush.txt};
        \end{axis}
    \end{tikzpicture}}
    \phantomsubcaption{}
    \label{fig:buffer_vs_aspect_ratio1}
  \end{minipage}%
    \caption{Effective frequency, \gls{RtWL}, \#DRCs, and number of inserted buffers for all considered floorplan styles as a function of the aspect ratio.}
    \label{fig:aspect_ratio_results}
\end{figure}
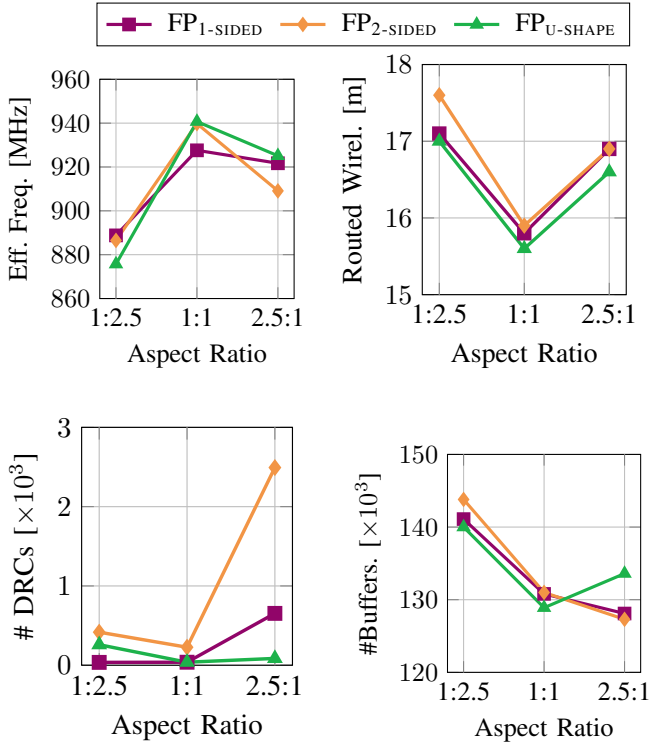

In this section, we analyze and compare the three floorplan styles as discussed in \cref{sec:impl_basefp} for three aspect ratios (width:height)---a tall (\aspectratio{1}{2.5}), a square (\aspectratio{1}{1}), and a wide cluster tile (\aspectratio{2.5}{1})---with a uniform cluster area of \SI{0.9}{\square\milli\meter}.
This aspect ratio selection enables a variety of hierarchical scale-out floorplans, \eg the ones shown in \cref{fig:fp_toplevel}.
Our findings are summarized in \cref{tab:aspect_ratio_results}.
In addition, we plot four key metrics, effective frequency, \gls{RtWL}, number of DRC violations, and the number of inserted buffers in \cref{fig:aspect_ratio_results}.
Overall, moving away from a square floorplan causes a noticeable \gls{QOR} degration.

\begin{table*}[t]
  \centering
  \caption{Physical implementation results of the cluster soft tile instances with aspect ratios ranging from \aspectratio{1}{2.5} to \aspectratio{2.5}{1} and base floorplans \FPone{}, \FPtwo{}, and \FPush{}.}
  \begin{tabular}[ht]{@{}rlllclllclll@{}}
    \toprule
    Aspect Ratio & \multicolumn{3}{c}{\aspectratio{1}{2.5}}  & & \multicolumn{3}{c}{\aspectratio{1}{1}}  & & \multicolumn{3}{c}{\aspectratio{2.5}{1}}\\
    \cmidrule{2-4}\cmidrule{6-8}\cmidrule{10-12}
    Floorplan & \FPone & \FPtwo & \FPush & & \FPone & \FPtwo & \FPush & & \FPone & \FPtwo & \FPush\\
    \midrule
    Eff. Freq. [\si{\mega\hertz}] & 888.8   & 886.5   & 875.7   & & 927.6   & 939.8   & 940.7   & & 921.7   & 909.1    & 925.1   \\
    \gls{TNS} [\si{\nano\second}] & -33.8   & -48.2   & -103.3  & & -25.5   & -30.2   & -24.7   & & -37.5   & -40.2    & -78.2   \\
    \#Violating Paths             & 5352    & 5787    & 6819    & & 4890    & 5372    & 4459    & & 6163    & 5871     & 8271    \\
    \gls{RtWL} [\si{\meter}]      & 17.1    & 17.6    & 17.0    & & 15.8    & 15.9    & 15.6    & & 16.9    & 16.9     & 16.6    \\
    \#DRCs                        & 36      & 417     & 259     & & 38      & 227     & 38      & & 654     & 2943     & 86      \\
    \#Buffers                     & \num{141.1e3}        & \num{143.8e3}        & \num{140.0e3}        & & \num{130.8e3}        & \num{131.0e3}        & \num{128.9e3}        & &
    \num{138.1e3}        & \num{137.3e3}         & \num{133.6e3} \\
    Cell Density                  & \SI{59.5}{\percent}  & \SI{60.7}{\percent}  & \SI{59.7}{\percent}  & & \SI{57.3}{\percent}  & \SI{57.9}{\percent}  & \SI{57.4}{\percent}  & & \SI{58.7}{\percent}  & \SI{58.9}{\percent}   & \SI{58.5}{\percent}  \\
    \bottomrule
  \end{tabular}
  \label{tab:aspect_ratio_results}
%   \vspace{0.1cm}
\end{table*}

\begin{figure*}[t]
  \begin{minipage}{.12\textwidth}
    \centering
    % \raggedright
    \includegraphics[scale=0.16]{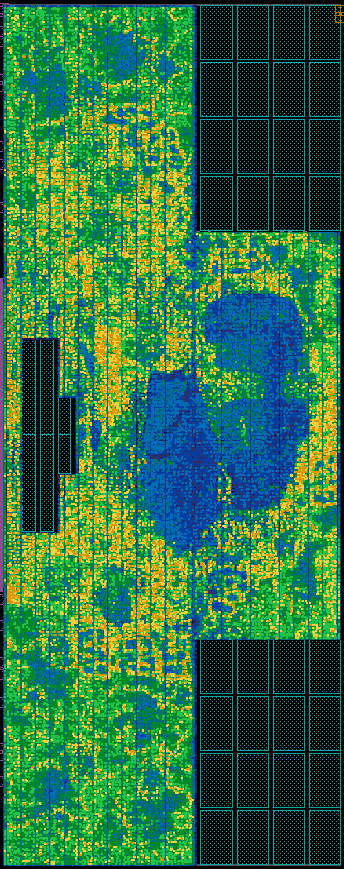}
    \subcaption{}
    % \subcaption{Narrow \FPone{}.}
    \label{fig:cell_density_oneside_Q2.5}
  \end{minipage}\hfill%
  \begin{minipage}{.12\textwidth}
    \centering
    % \raggedright
    \includegraphics[scale=0.16]{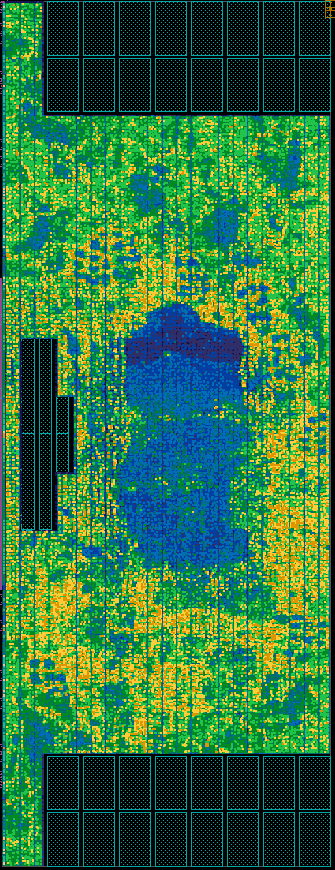}
    \subcaption{}
    % \subcaption{Narrow \FPtwo{}.}
    \label{fig:cell_density_twoside_Q2.5}
  \end{minipage}\hfill%
  \begin{minipage}{.12\textwidth}
    \centering
    % \raggedright
    \includegraphics[scale=0.16]{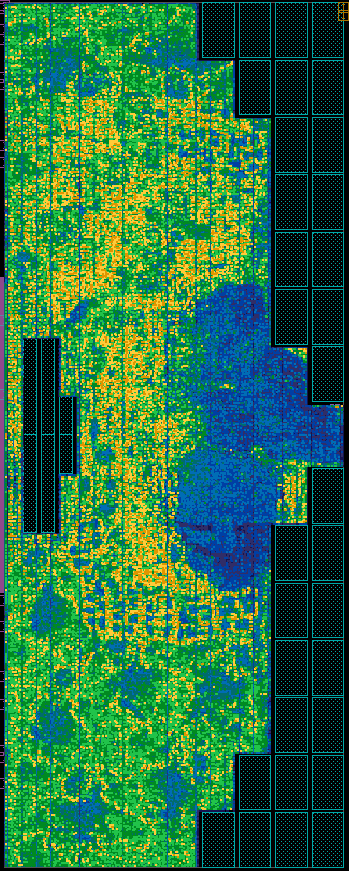}
    \subcaption{}
    % \subcaption{Narrow \FPush{}.}
    \label{fig:cell_density_ushape_Q2.5}
  \end{minipage}\hfill%
  \begin{minipage}{.60\textwidth}
    \centering
      \begin{minipage}{0.61\textwidth}
        \centering
        \vspace{0.47cm}
        \includegraphics[scale=0.14]{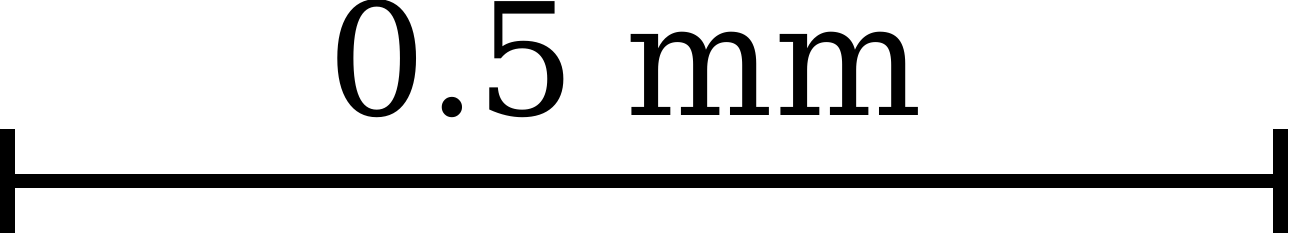}
        \vspace{0.47cm}
      \end{minipage} \hfill%
      \begin{minipage}{0.38\textwidth}
        \raggedleft
        \centering
        \vspace{0.47cm}
        \includegraphics[height=0.9cm]{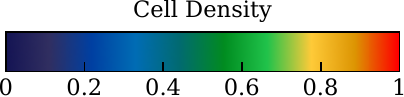}
        \vspace{0.47cm}
      \end{minipage}
      \begin{minipage}{.32\textwidth}
        \centering
        % \raggedleft
        \includegraphics[scale=0.12]{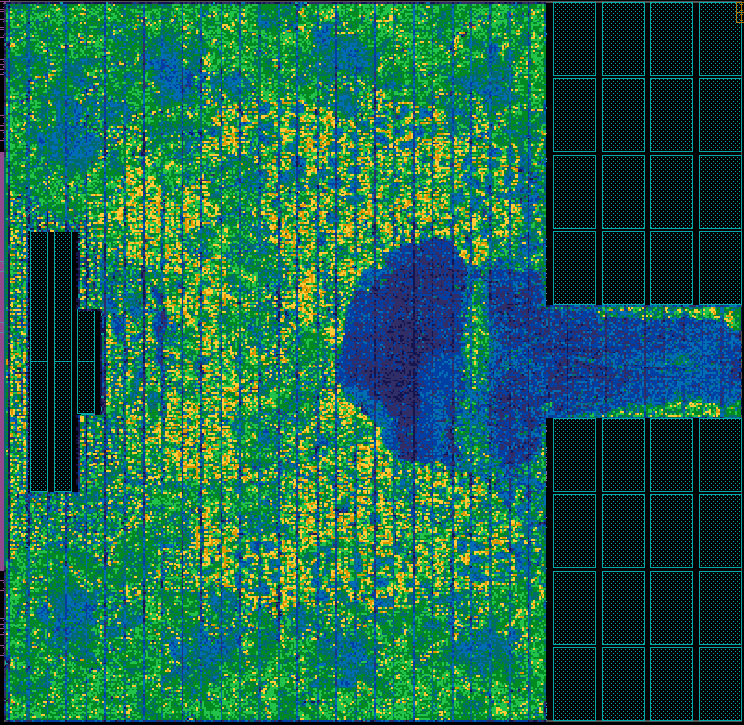}
        \subcaption{}
        % \subcaption{Square \FPone{}.}
        \label{fig:cell_density_oneside_Q1.0}
      \end{minipage}\hfill%
      \begin{minipage}{.32\textwidth}
        \centering
        % \raggedleft
        \includegraphics[scale=0.12]{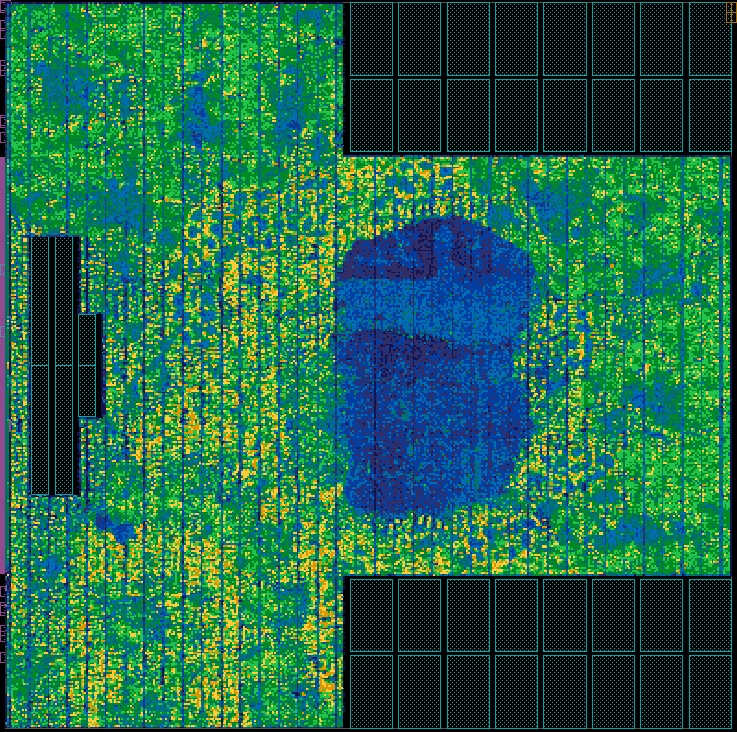}
        \subcaption{}
        % \subcaption{Square \FPtwo{}.}
        \label{fig:cell_density_twoside_Q1.0}
      \end{minipage}\hfill%
      \begin{minipage}{.32\textwidth}
        \centering
        % \raggedleft
        \includegraphics[scale=0.12]{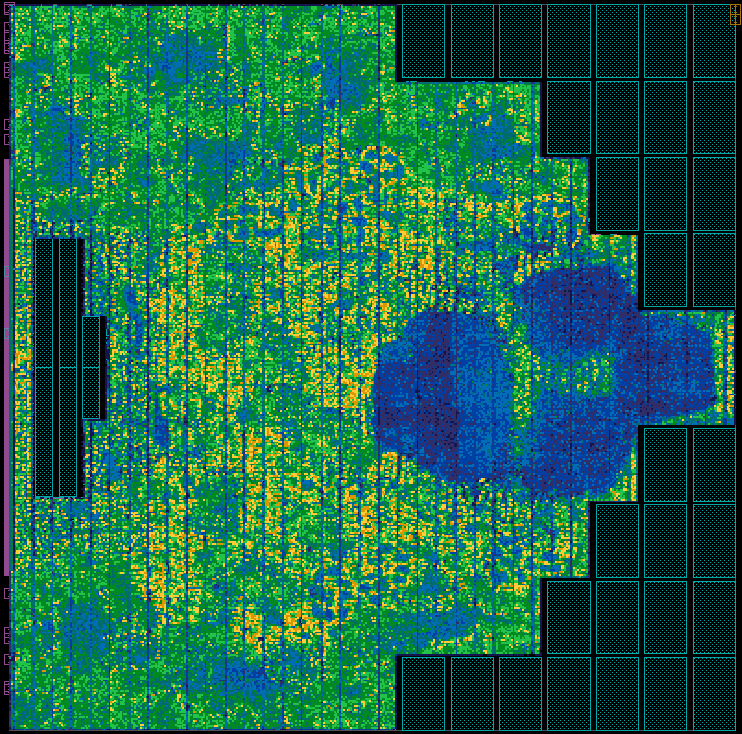}
        \subcaption{}
        % \subcaption{Square \FPush{}.}
        \label{fig:cell_density_ushape_Q1.0}
      \end{minipage}%
    \end{minipage}%
  \vspace{1ex}
  \\
  \begin{minipage}{.32\textwidth}
    \centering
    \raggedright
    \includegraphics[scale=0.10]{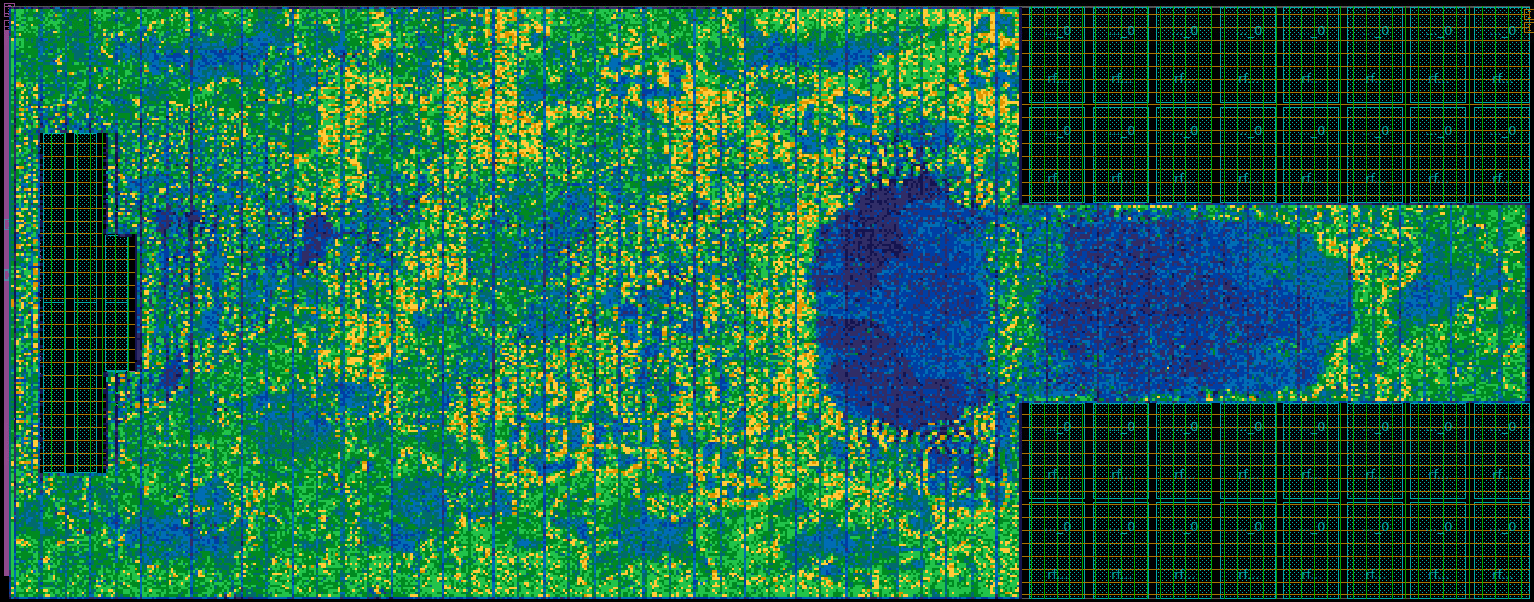}
    \subcaption{}
    % \subcaption{Wide \FPone{}.}
    \label{fig:cell_density_oneside_Q0.4_wChannel}
  \end{minipage}\hfill%
  \begin{minipage}{.32\textwidth}
    \centering
    \includegraphics[scale=0.10]{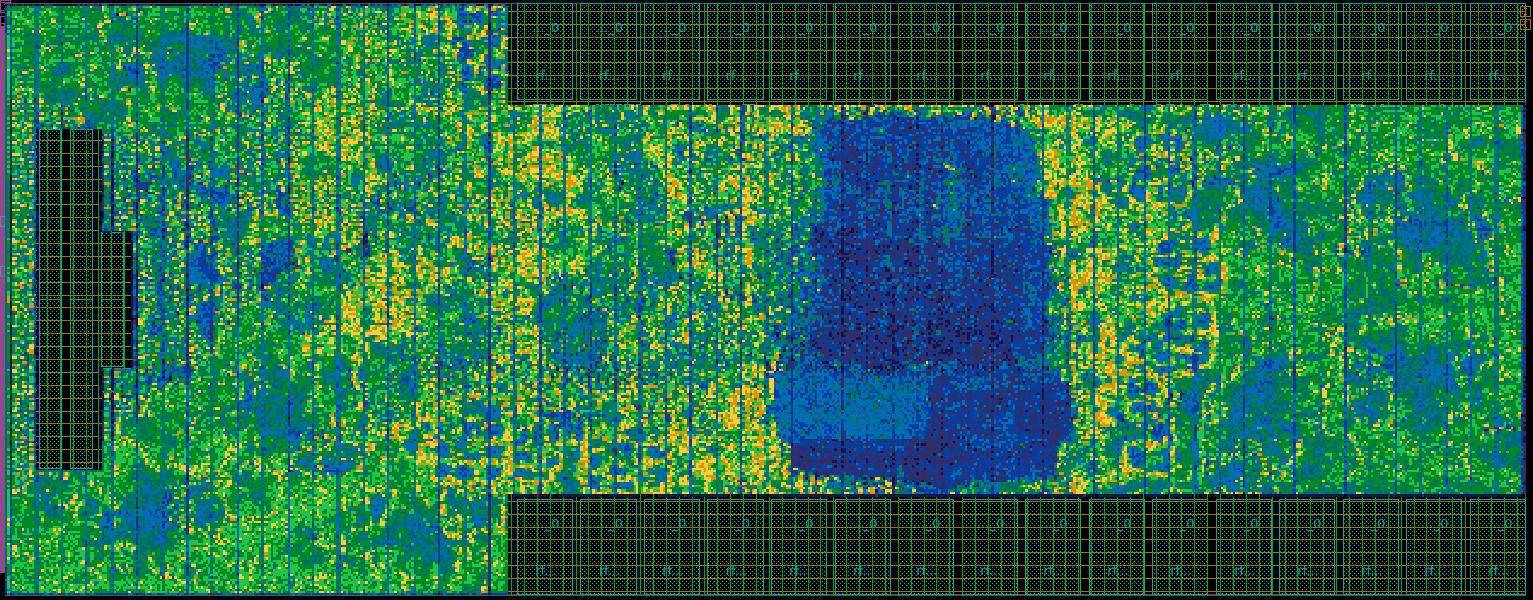}
    \subcaption{}
    % \subcaption{Wide \FPtwo{}.}
    \label{fig:cell_density_twoside_Q0.4_wChannel}
  \end{minipage}\hfill%
  \begin{minipage}{.32\textwidth}
    \centering
    \raggedleft
    \includegraphics[scale=0.10]{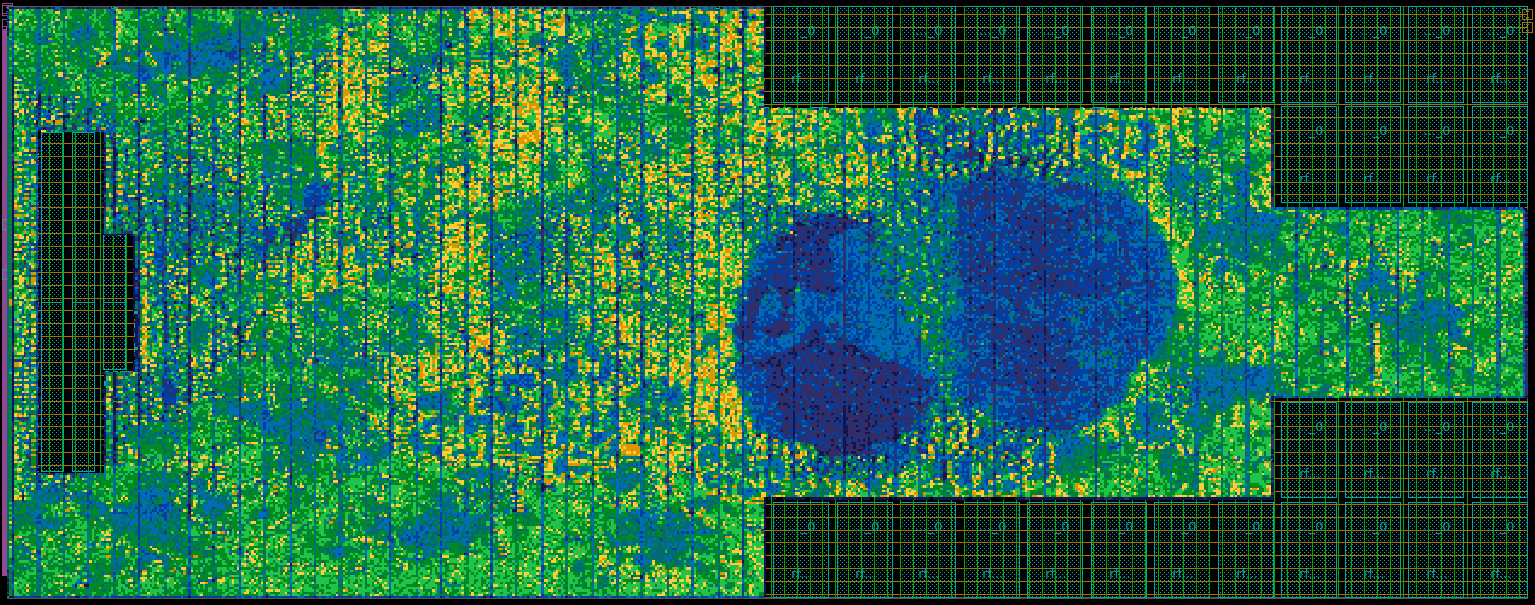}
    \subcaption{}
    % \subcaption{Wide \FPush{}.}
    \label{fig:cell_density_ushape_Q0.4}
  \end{minipage}%
  \caption{Standard cell density map, from dark blue to red, for each considered aspect ratio: \aspectratio{1}{2.5} (a), (b) (c), \aspectratio{1}{1} (d), (e), (f), and \aspectratio{2.5}{1} (g), (h), (i). Within each aspect ratio, the images represent the macro placement styles \FPone{}, \FPtwo{}, and \FPush{}, respectively. The black regions contain the \gls{SRAM} macros. Images to scale.}
  \label{fig:cell_density}
\end{figure*}

For our three floorplan styles, pushing the floorplan to a tall aspect ratio of \aspectratio{1}{2.5}, shown in \cref{fig:cell_density_oneside_Q2.5,fig:cell_density_twoside_Q2.5,fig:cell_density_ushape_Q2.5}, degrades the overall \gls{QOR} more than pushing to a wide aspect ratio of \aspectratio{2.5}{1}, shown in \cref{fig:cell_density_oneside_Q0.4_wChannel,fig:cell_density_twoside_Q0.4_wChannel,fig:cell_density_ushape_Q0.4}.
Similarly to the square floorplan \FPtwo{} in  \cref{fig:cell_density_twoside_Q1.0}, the three floorplans \FPone{}, \FPtwo{}, and \FPush{} for the tall aspect ratio of \aspectratio{1}{2.5} in \cref{fig:cell_density_oneside_Q2.5,fig:cell_density_twoside_Q2.5,fig:cell_density_ushape_Q2.5} have an asymmetric module placement: the \glspl{CC} cannot be placed in the central location between the interconnect and the \gls{IDol} banks and are squeezed to one side.
This placement impacts their \gls{QOR}.
Out of those three instances, \FPtwo{} and \FPush{} are unfeasible due to a high DRC violation count.
Only \FPone{} is feasible, although at an operating frequency \SI{5.5}{\percent} lower than the fastest square-shaped instance.

% Similarly, to the square (\aspectratio{1}{1}) floorplan \FPtwo{} in  \cref{fig:cell_density_twoside_Q1.0}, the two floorplans \FPtwo{} and \FPush{} for the tall aspect ratio of \aspectratio{1}{2.5} in \cref{fig:cell_density_twoside_Q2.5,fig:cell_density_ushape_Q2.5} are asymmetric: the \gls{MulDiv}, and, therefore, also the \glspl{CC}, cannot be placed in the central location close to the \gls{IDol} banks and are squeezed to the side.
% With 417 and 259 DRC violations the two floorplans are infeasible.
% Thus, \FPone{} is with 36 DRC violations and a \gls{TNS} of \SI{-33.8}{\nano\second} the only feasible floorplan for the tall aspect ratio.
% The effective frequency is with \SI{888.8}{\mega\hertz} \SI{4.2}{\percent} worse than the square \FPone{} floorplan and \SI{5.5}{\percent} worse than the square \FPush{} floorplan.

\Cref{fig:cell_density_oneside_Q0.4_wChannel,fig:cell_density_twoside_Q0.4_wChannel,fig:cell_density_ushape_Q0.4} show the cell density maps for \FPone{}, \FPtwo{}, and \FPush{} for the wide aspect ratio of \aspectratio{2.5}{1}.
The smaller height for these floorplans requires the \gls{SPM} banks to be arranged in more columns causing the fully-connected crossbar (blue region) to be placed more towards the center of the floorplan. The resulting free area on the right side gets filled with \glspl{FPU}. For both \FPone{} and \FPtwo{}, the fully-connected crossbar is squeezed and stretched by the wide \gls{SPM} bank rows, respectively. Thus, all connections between the \glspl{CC} and the \gls{FPU} must be routed through the already congested crossbar area.
The lack of routing resources in that area increases the number of DRC violations for \FPone{} and \FPtwo{}.
In contrast, the arrangement of the \gls{SPM} banks in the \FPush{} instance leads to an undisturbed crossbar placement.
This allows the \glspl{CC} to \gls{FPU} connections to route around the high routing congestion crossbar area.
Nevertheless, the overall feasibility of \FPush{} with 86 DRC violations remains questionable.

% We conclude, targeting more extreme aspect ratios for a design might require dedicated floorplan evaluations. While more extreme aspect ratios can be achieved, they usually come with an overall degradation of \gls{QOR}.

%%%%%%%%%%%%%%%%%%%%%%%%%%%%%%%%%%%%%%%%%%%%%%%%%%%%%%%%%%%%%%%%%%%%%%%%%
\subsection{Hierarchical Design Flow Recommendations}

Our analysis shows that the \gls{QOR} of a soft tile is particularly sensitive to aspect ratio variations and highly depends on the floorplan style, \ie on the memory macro placement.
As the main compute unit, the latency-critical cluster should strive to achieve the best possible \gls{QOR}.
The higher hierarchy levels use latency-tolerant and pipelineable interconnects which can easily be tuned to match the cluster's performance.
However, additional requirements might propagate down the hierarchy when designing the top-level floorplan.
For example, the top-level floorplans from \cref{fig:fp_onesided,fig:fp_ushape} might require larger channels between the bottom row of quadrants to allow the global crossbar to route to a die-to-die interface \gls{IP} or an \gls{HBM} PHY, requiring a non-square aspect ratio for the cluster tile.
Similarly, the narrow channels between the three rows of quadrants in \cref{fig:fp_twosided} might lead to congested regions, which implies different aspect ratio requirements for the cluster tile.

Based on our findings, we propose to follow our hierarchical implementation approach:
in a first step, the designer should evaluate various cluster-level floorplan styles to explore a set of aspect ratios which enable a variety of top-level floorplans.
The knowledge of the achievable cluster-level \gls{QOR} and optimal floorplans for each aspect ratio then allows the further exploration of a set of top-level floorplans using only soft tile shapes that meet the desired \gls{QOR}.

%%%%%%%%%%%%%%%%%%%%%%%%%%%%%%%%%%%%%%%%%%%%%%%%%%%%%%%%%%%%%%%%%%%%%%%%%
\section{Conclusions}
\label{sec:conclusion}

Typically, modern high-performance accelerator architectures are based on tightly-coupled clusters of \glspl{PE}, physically implemented as rectangular soft tiles.
These tiles are often highly replicated and interconnected with a latency-tolerant \gls{NOC} to build scalable high-performance computing systems.
The tile's size and aspect ratio strongly impact the achievable operating frequency and energy efficiency.
Nevertheless, they should be as flexible as possible to enable a high density solution for the complete design.
In this paper, we focus on an open-source, high-performance multicore cluster as a soft tile, which can be used to build a high-performance many-cluster system by cluster replication.
Based on architectural analysis, we proposed three floorplan styles which can systematically be adapted to different aspect ratios.
% Based on macro placement guidelines extracted from our cluster architecture, we proposed three floorplan styles which can be easily be adapted to different aspect ratios.

We then explored the \gls{QOR} of placed-and-routed cluster implementations as a soft tile in the \textsc{GlobalFoundries} \SI{12}{\nano\meter} advanced FinFET technology node, considering wide (\aspectratio{2.5}{1}), square, and tall (\aspectratio{1}{2.5}) cluster aspect ratios with all proposed floorplan styles.
No single floorplan style leads to good \gls{QOR} across all aspect ratios.
\FPone{} is to be used when higher-level tiles require a tall cluster floorplan, and \FPush{} is better suited when a wide cluster tile is required.
Considering the overall \gls{QOR}, the best \aspectratio{2.5}{1} instance achieves an effective frequency \SI{1.7}{\percent} lower than the best square instance, although the large \gls{TNS} and DRC violation count poses doubts on its feasibility.
On the clusters with an aspect ratio \aspectratio{1}{2.5}, only the \FPone{} instance was feasible, albeit at an operating frequency \SI{5.5}{\percent} lower than the best square cluster instance.

Overall, the results suggest that pre-characterizing the building block's \gls{QOR} for different aspect ratios and floorplan styles, combined with a preliminary investigation of the top-level layout, can help the designer find top-level hierarchical floorplans that are feasible and achieve high performance.

\bibliographystyle{IEEEtran}
\bibliography{IEEEabrv,bib/base,bib/pauling}

\end{document}